\documentstyle[twoside,fleqn,npb,epsfig]{article}
\newcommand{\AmS}{{\protect\the\textfont2
  A\kern-.1667em\lower.5ex\hbox{M}\kern-.125emS}}

\title{$\bar u - \bar d$ asymmetry - a few remarks.}

\author{A.~Szczurek\address{H.~Niewodniczanski Institute of Nuclear Physics,
ul.~Radzikowskiego 152, Krak\'ow, Poland}, V.~Uleshchenko$^{\rm a}$}

\begin{document}

\begin{abstract}
We make a few remarks on possible sources of uncertainties
of the $\bar d - \bar u$ asymmetry obtained by different
methods and comment on its possible verification
in the future. In addition we comment on its present
understanding.
\end{abstract}


\maketitle

In the last year both the E866 collaboration at Fermilab
(Drell-Yan production of dimuons) and HERMES collaboration at DESY
(semi-inclusive production of charged pions) published their new results
on $\bar d - \bar u$ asymmetry in the nucleon \cite{E866,Hermes}.
During the DIS99 conference both groups have presented their new
results with somewhat better statistics \cite{E866_a,Hermes_a}.
The new results complement the older results obtained by the NMC
on the Gottfried integral \cite{NMC} and earlier Drell-Yan experiment
NA51 from CERN \cite{NA51}.

The E866 collaboration measured the ratio of the cross sections:
$ \sigma_{pd}^{DY} / \sigma_{pp}^{DY} $. This ratio is extremely
sensitive to the $\bar d / \bar u$ ratio. The $\bar d / \bar u$
ratio is extracted in an iterative procedure assuming leading order
formulae and that valence quark distributions as well as
$\bar u + \bar d$ are as given by PDF's \cite{PDF}.
Next the difference $\bar d - \bar u$ is obtained from
\begin{equation}
\bar d - \bar u =
\frac{\bar d / \bar u - 1}{\bar d / \bar u + 1}
\cdot [ \bar u + \bar d ] \; .
\end{equation}
In practice the E866 collaboration uses $\bar u + \bar d$ from
one of the global NLO fit to the world data. Here different global
PDF fits \cite{PDF} yield roughly similar result for $ x > 0.05$.
In this range of $x$ the sum is strongly constrained from the
(anti)neutrino experiments. At smaller values of $x$ one should worry
about the consistency of using NLO PDF's in LO formulae. At larger values
of $x > 0.4$ our knowledge of $\bar u + \bar d$ is rather limited.
This must be taken into account particularly seriously
in the planned P906 experiment \cite{P906}.
The average value of $Q^2$ in the E866 experiment is high enough
not to expect any higher-twist effects.

In obtaining the $\bar d - \bar u$ asymmetry the HERMES collaboration
assumes the factorization between the hard scattering process and
the hadronization of the struck quark
\begin{equation}
N^{\pi^{\pm}}(x,z) \propto \sum_i e_i^2
[q_i(x) D_{q_i}^{\pi^{\pm}}(z) +
\bar q_i(x) D_{q_i}^{\pi^{\pm}}(z) ] \; ,
\end{equation}
i.e. assumes implicitly the validity of the parton model.
The isospin symmetry (IS) between proton and neutron reduces the number
of light-quark fragmentation functions to two, favoured
and disfavored. Then \cite{Hermes}
\begin{equation}
\frac{1+r}{1-r} = \frac{u-d+\bar u - \bar d}{[u-\bar u] - [d-\bar d]}
J(z) \; ,
\end{equation}
where $r(x,z) = \frac{N_p^{\pi^-}(x,z) - N_n^{\pi^-}(x,z)}
                     {N_p^{\pi^+}(x,z) - N_n^{\pi^+}(x,z)}$
and $J(z) = \frac{3}{5} ( \frac{1+D'(z)}{1-D'(z)} )$,
$D'(z) = D_u^{\pi^-}(z) / D_u^{\pi^+}(z)$.

The HERMES experiment is a fixed target experiment with the beam
electron energy of about 30 GeV, i.e. the small $x$ is associated
with relatively small $Q^2$. In the lowest $x$ bin the average $Q^2$
is only slightly larger than 1 GeV$^2$. It is an open problem how big
are the higher-twist effects beyond the parton model at such small
values of $Q^2$. A simple estimate of the VDM contribution to
the structure function shows that it can be of the order of 20 \%.
Assuming IS for hadronic components we get for nucleon - virtual vector
meson scattering:
$\sigma(p V^0 \rightarrow \pi^+) = \sigma(n V^0 \rightarrow \pi^-) >
 \sigma(p V^0 \rightarrow \pi^-) = \sigma(n V^0 \rightarrow \pi^+)$.
The inequality comes from the fact that presumably $u_p > d_p$ and
$\bar d_p > \bar u_p$.
Thus the presence of hadronic component would lead
to a reduction of the experimentally extracted quantity $r(x,z)$.
This means that the corresponding purely partonic quantity
(exclusively theoretical quantity) would be bigger.
This would result in a smaller $\bar d - \bar u$.
No quantitative estimate of the effect has been made up to now.

Assuming the validity of the parton model the HERMES collaboration
extracts the quantity $ (\bar d - \bar u) / (u - d) $. The denominator
is in our opinion not extremely well known. The measured region of $x$
is sensitive to the meson cloud effects \cite{traditional_MC}.
We wish to stress a poorly known fact that the meson cloud effects
contribute both to the sea and valence quark distributions.
Therefore it is not clear whether the PDF parametric forms
used in global fits (even for valence quark distributions) are flexible
enough to accomodate those effects.

Both the E866 and HERMES collaborations tried to estimate the integral
$\int_0^1 [\bar d - \bar u ] \; dx$. It appears that the number obtained
by the E866 collaboration is slightly lower than those obtained by the
NMC and HERMES collaborations. Is it a random statistical fluctuation
or there is a physical reason behind it?
Recently we have shown \cite{SU}
that a two component model which includes the VDM contribution (modified
for large $x$ for finite fluctuation times of the hadronic component of
the photon) and a modified partonic component (vanishing at small $Q^2$
\footnote{the traditional parton model do not posses this property})
can describe both the proton and deuteron structure functions in the
broad range of $x$ and $Q^2$; considerably better than
the pure QCD-improved parton model.
The model from \cite{SU} has interesting predictions
for $F_2^p - F_2^n$. Here the VDM contribution cancels and one
is left with a modified partonic component which tends to zero at
$Q^2 \rightarrow$ 0. Already at $Q^2$ as large as 4 GeV$^2$
(typical for NMC data) we find \cite{SU} a non-negligible reduction of
the parton model result.
This prediction for a strong $Q^2$ dependence of the $F_2^p - F_2^n$
seems to be confirmed by the world data for $F_2^p$ and $F_2^d$.
The $F_2^p - F_2^n$ as a relatively small quantity is very sensitive
to statistical uncertainties and cannot be obtained by a simple
subtraction of $F_2^p - (F_2^d - F_2^p)$. Here one can use the method
proposed by the NMC \cite{NMC}. The QCD improved parton model seems
to fail for the extracted $F_2^p - F_2^n$ already at $Q^2$ as large
as 7 GeV$^2$ \cite{SU_GS}. In the language of the higher-twist expansion
this means that the twist-4 contribution is rather large and negative.
This phenomenological observation is consistent with a recent QCD
lattice result \cite{QCD_lattice}. The substantial higher-twist effects
strongly modify our present understanding of the applicability of
the QCD-improved parton model.
The strong $Q^2$ dependence of the $F_2^p - F_2^n$
can potentially explain the difference between the E866 (large $Q^2$) and
NMC (small $Q^2$) results.

Despite the not fully resolved problems, mentioned above,
the new experiments provided valuable information on
$\bar d - \bar u$ asymmetry in the nucleon and constitute
a useful input which can be used to constrain PDF's.
The LO Altarelli-Parisi evolution equations generate an equal number
of $\bar u - u$ and $\bar d - d$ pairs. The two-loop evolution
gives a rather negligible effect \cite{RS79}. Because perturbative
QCD is not able to explain the large asymmetry and
the Gottfried Sum Rule violation it is clear that the relevant
physics must be of nonperturbative origin.

The chiral symmetry and chiral symmetry breaking leads to the presence
of the pion cloud in the nucleon. This concept provides
the most natural and economic explanation of the asymmetry
(see for instance \cite{traditional_MC}).
There exists 2 technical formulations of such a model.
In the traditional nuclear physics formulation the physical nucleon
is expanded in terms of the meson-baryon Fock states as
\begin{equation}
| N > = |N_0> + | \pi N' > + | \pi \Delta > + etc.
\end{equation}
The most complete version of the model has been presented in
\cite{traditional_MC}.
If the coupling constants are fixed from low-energy hadronic physics
and the vertex form factors from high-energy production of barions,
the model leads to (a) $\pi^+ > \pi^0 > \pi^-$, (b) the number of
pions in the nucleon $N(\pi)$ = 0.2 - 0.3, and
(c) the pion distribution $P(x_{\pi})$ which peaks at $x_{\pi}
\sim$ 0.2 - 0.3.
The latter means that the momentum fraction of the neutron associated
with the pion would be about 0.7 - 0.8. This is fully consistent
with the spectra of leading neutrons at HERA \cite{LN}.

Parallel to the traditional approach, the effective chiral quark theory
provides an alternative explanation. Here the relevant degrees of
freedom are constituent quarks and Goldstone bosons. The most
extended analysis of the light-antiquark asymmetry in this type of
models can be found in \cite{chiral_MC}. If the constituent quark -
pion vertex form factor is fixed to the size of the Gottfried
Sum Rule violation then: (a) $\pi^+ : \pi^0 : \pi^-$
= 2:3/2:1, (b) $N(\pi)$ = 0.6 - 0.7 and (c) $P(x_{\pi})$ which
peaks at $x_{\pi} \sim$ 0.1.

The recent E866 experiment at Fermilab has reported a first
high-precision mapping of the x-dependence of the $\bar u - \bar d$
asymmetry with the finding that the difference of
$\bar d - \bar u$ seems to vanish at large $x \ge$ 0.3.
This surprising observation was not predicted by
models which used only limitations on leading baryons.
It was shown in \cite{NSSS98} that if the information on leading
pions in hadronic reactions is used in addition, to limit
the hadronic vertex form factors, than the new E866 data on
$\bar d - \bar u$ can be described automatically.

In all the present experimental analyses: muon deep inelastic scattering
\cite{NMC}, E866 Drell-Yan experiment \cite{E866} and HERMES
semi-inclusive pion production \cite{Hermes} both the proton and neutron
(deuteron) targets are used.
In order to obtain the information on the $\bar u - \bar d$ asymmetry
one assumes IS of quark (antiquark) distributions in the proton and neutron i.e.
\begin{eqnarray}
&&u_n(x) = d_p(x), \;\;\; d_n(x) = u_p(x), \;\;\; \nonumber \\
&&\bar u_n(x) = \bar d_p(x), \;\;\; \bar d_n(x) = \bar u_p(x) \; .
\end{eqnarray}
Such a symmetry for PDF was never tested experimentally.
Recently a simple analysis of the muon and neutrino structure function
led to the conclusion of substantial isospin violation
in PDF's \cite{BLT98}.
Ascribing all the observed effect to isospin violation is rather an
extreme view \cite{Bodek}. Even if the true violation of IS
is much smaller than suggested in \cite{BLT98} it remains
essentially unknown experimentally.
Therefore all the present analyses are to some extent biased by
the explicit assumption of IS.
In \cite{SUHS97} we have suggested how to test the $\bar u - \bar d$
asymmetry avoiding the assumption of IS.
We suggested to measure at RHIC the asymmetry
\begin{equation}
A(p p \rightarrow W^{\pm}) =
\frac{\sigma(p p \rightarrow W^{+}) -
      \sigma(p p \rightarrow W^{-})}
     {\sigma(p p \rightarrow W^{+}) +
      \sigma(p p \rightarrow W^{-})}
\end{equation}
as a function of $W$-boson rapidity or similar asymmetry for charged
leptons from the decay of $W$ bosons. The $x$-dependence of the asymmetry
could be obtained by varying the beam energy at RHIC.

\end{document}